\documentclass[twocolumn,superscriptaddress,showpacs,preprintnumbers,amsmath,amssymb]{revtex4}

\newcommand{\bb}{\begin{equation}}
\newcommand{\en}{\end{equation}}
\usepackage{graphicx}
\usepackage{color}

\begin{document}

\draft

\title{Thermal expansion within a chain of magnetic colloidal particles}

\author{D. Lacoste}
\affiliation{Physico-Chimie Th\'eorique,
UMR 7083 CNRS-ESPCI, 10 rue Vauquelin, 75231 Paris Cedex 05, France}
\author{C. Brangbour}
\affiliation{Laboratoire de Collo\"{\i}des et mat\'eriaux
divis\'es, ESPCI ParisTech, UPMC, UMR 7612 CNRS, 10 rue Vauquelin, 75005 Paris, France}
\author{J. Bibette}
\affiliation{Laboratoire de Collo\"{\i}des et mat\'eriaux
divis\'es, ESPCI ParisTech, UPMC, UMR 7612 CNRS, 10 rue Vauquelin, 75005 Paris, France}
\author{J. Baudry}
\affiliation{Laboratoire de Collo\"{\i}des et mat\'eriaux
divis\'es, ESPCI ParisTech, UPMC, UMR 7612 CNRS, 10 rue Vauquelin, 75005 Paris, France}

\begin{abstract}
We study the thermal expansion of chains formed by self-assembly
of magnetic colloidal particles in a magnetic field. Using
video-microscopy, complete positional data of all the particles of
the chains is obtained. By changing the ionic strength of the
solution and the applied magnetic field, the interaction potential
can be tuned. We analyze the thermal expansion of the chain using
a simple model of a one dimensional anharmonic crystal of finite
size.
\end{abstract}

\date{\today}
 \maketitle

It is well-known that the expansion of a solid with temperature is
associated with the anharmonicity of the interaction between the
particles of the solid. This was shown with standard solids like
lead, using X-ray absorption measurements \cite{stern}, and in
this reference, the thermal expansion is analyzed using a
one-dimensional anharmonic oscillator model. This suggests that it
would be valuable to carry out such a study with a one-dimensional
crystal. Although no true long range order can exist in
one-dimensional systems in the thermodynamic limit, many systems
are one-dimensional phases of finite length. A classical example
is the mercury salts \cite{tom}, but more recently, carbon
nanotubes, phases of gases adsorbed inside carbon nanotubes
\cite{cole}, or the B phase of DNA have attracted considerable
interest due to their one dimensional character.

In these one-dimensional crystals, only few experiments have
investigated the thermal expansion. One reason for that is that
these systems do not allow for an easy control of the interaction
between th
e particles (or the temperature). With colloidal
systems, and in particular magnetic colloids, it is easy to obtain
low dimensional condensed phases in which the interaction between
the particles can be tuned.
  This strategy was already used in \cite{maret}, where magnetic
  colloids were used in a beautiful
  experimental investigation of the 2D melting transition.
To our knowledge, no comparable experimental study of 1D colloidal
crystals has been carried out.

In the present letter, video microscopy is used to study the
thermal expansion in a chain of magnetic colloids. At high
magnetic field, such a chain can be
viewed as a quasi-one-dimensional condensed phase of colloidal
particles. In this system, the interaction potential
can be determined experimentally from the dependence of the
average interparticle distance with the
applied magnetic field \cite{remi,leal94}.
We analyze quantitatively these experiments using
Monte-Carlo simulations. As in standard solids, we find that the
expansion as function of temperature is controlled by the
anharmonicity of the interaction potential. In magnetic colloids,
the temperature is measured by the inverse of a dimensionless
number, $\lambda$ which measures the strength of the dipolar
interaction with respect to the thermal energy. The anharmonicity
of the interaction potential is varied experimentally by changing
the ionic strength of the solution, which controls the
electrostatic repulsive part of the potential.


In this paper, we describe a colloidal chain as a 1D lattice of
$N$ colloidal particles of diameter $d$ with a nearest-neighbor
interaction potential $u(r)$, where $r$ is the spacing between two
neighboring particles in the chain. Let us recall some well-known
general features of this system before coming to the specific case
of magnetic colloids. Calculations are conveniently carried out in
the ensemble where the temperature $T$ and the volume $L$ (which
corresponds to the length of the chain in 1D) are constant. The
pressure $P$ (which corresponds to a force in 1D) is the Lagrange
multiplier associated with the constraint of constant volume.
Thanks to the nearest-neighbors assumption, the partition function
of this problem $Z$ can be calculated exactly \cite{ma}. It is
$Z=\xi^{N+1}/\Lambda^N$, where $\beta=1/k_B T$,
\begin{equation}
\xi=\int dr e^{-\beta \left( u(r)+Pr \right)},
\end{equation} and $\Lambda$ is the de Broglie
length. Thus, the enthalpy $G$ equals $N k_B T \log (\Lambda/\xi)$
(neglecting $1/N$ corrections). At low temperatures, this system
behaves as a solid, which suggests to expand the potential $u(r)$
about its minimum obtained at $r_{min}$. When the function
$v(r,r_{min})$ defined as $u(r)-u(r_{min})$, is symmetric with
respect to $r_{min}$ (by that we mean precisely
$v(r,r_{min})=v(2r_{min}-r,r_{min})$), we find that there is no
thermal expansion $\langle r \rangle \simeq r_{min}$. At finite
temperature, no such symmetry is present: the average of the
position $\langle r \rangle$ differs from the most probable value
of the position $r_{min}$, which leads in general to thermal
expansion. This can be seen using a Taylor expansion of the form
\bb v(r,r_{min}) = \frac{1}{2}kx^2+gx^3, \en limited to cubic
order with $x=r-r_{min}$. After calculating $\xi$ by the saddle
point approximation, one obtains the length $L$ of the chain as a
derivative of the enthalpy   \cite{ma}
\begin{equation}
L=\frac{\partial G}{\partial P}=N \left( r_{min}-\frac{P}{k} +
\frac{3gP^2}{k^3}-\frac{3g k_B T}{k^2} \right).
\end{equation}
The fact that $L$ scales with $N$ indicates that the expansion is
uniform within the chain. Since we are interested in free chains,
the pressure $P=0$, and the average expansion per particle is
\cite{stern,keller}
\begin{equation} \label{expansion}
\langle x \rangle=\frac{L}{N}-r_{min}=-\frac{3g k_B T}{k^2}.
\end{equation}
This equation confirms that the expansion of a solid as a function
of temperature requires anharmonic terms such as the cubic term
proportional to $g$.
It is valid at the lowest (linear) order in $g$ or in $T$.
At the same order of perturbation, $\langle
(x-\langle x \rangle )^2 \rangle \simeq k_B T/k$, which is the
same expression as that obtained for an harmonic potential using
the equipartition theorem.

\begin{figure}
{\par {{\includegraphics[scale=1]{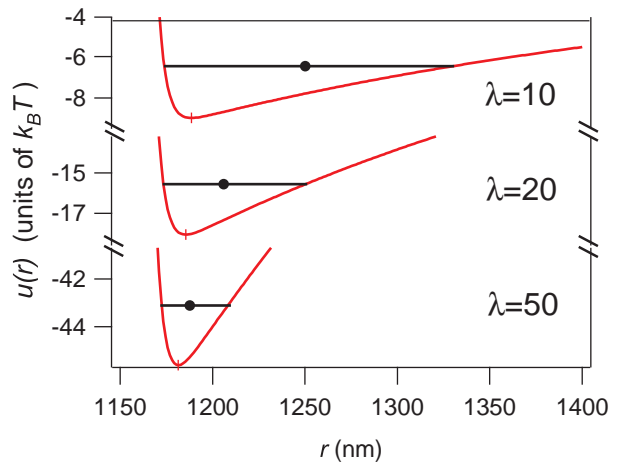}} }   
\par}
\caption{Total interaction potential $u(r)$ in units of $k_B T$ as function of the distance
$r$ between the centers of two neighboring colloidal particles. The curves correspond to different values of
$\lambda=10,20$ or 50 as shown. The black dot represents the average position $\langle r \rangle$, which is different from the minimum of the potential shown as a vertical stroke.
The particle diameter is
$d=1150$nm, the
zeta potential is $\Psi_0=-35$mV, and the ionic strength is 5mM, corresponding to a Debye length of about 4nm.} \label{Fig:potential}
\end{figure}

In the specific case of our experimental system of magnetic
colloidal chains, the dipolar magnetic part of the interaction
potential is attractive and of the form \bb \beta u_m(r)=-
\frac{\lambda d^3}{r^3}, \en where $\lambda=\beta m^2 \mu_0/2 \pi
d^3$,
 $m$ is the dipole strength and $\mu_0$ is the permeability of vacuum \cite{david,toussaint}.
In experiments, $\lambda$ is in the range of 10-1000. The
non-magnetic part of the pair potential is repulsive and can be
described as the sum of an hard sphere potential $u_{HS}(r)$
(which is zero for $r>d$ and infinite for $r=d$), and a
short-ranged electrostatic potential $u_{el}(r)$. In this paper,
we are only interested in the regime where the Debye length
$\kappa^{-1}$ is much smaller than the particle size. In this case
\begin{equation}\label{elect2}
u_{el}(r)= \pi \epsilon \Psi_0^2 d \ln \left( 1+e^{-\kappa (r -
d)} \right),
\end{equation}
in terms of the zeta potential $\Psi_0$, which is assumed to be
uniform and sufficiently low for the Debye theory to hold
\cite{hunter}. The total pair potential $u(r)$ is \bb
u(r)=u_{HS}(r)+u_{el}(r)+u_m(r). \en The function $u(r)$ is shown
as a function of $r$ on figure \ref{Fig:potential} for different
values of $\lambda$. An important feature of this potential is
that it is asymmetric for low values of $\lambda$.

\begin{figure}
{\par { \rotatebox{0}{\includegraphics[scale=1]{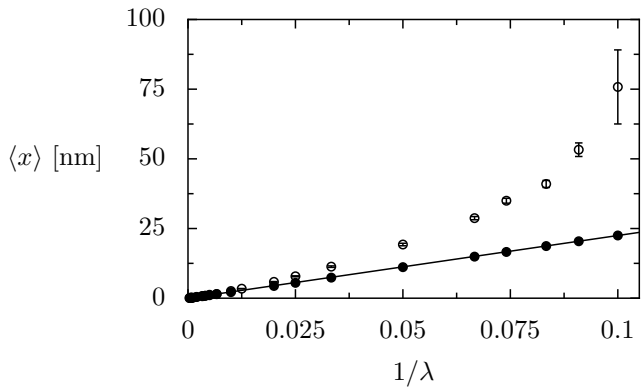}} }
\par}
\caption{Average expansion $\langle x \rangle$ versus effective temperature $1/\lambda$, the
empty circles are the simulation data points with the full potential and the filled circles
corresponds to the approximation of Eq.~\ref{expansion} with a cubic potential. The particle diameter, zeta potential and ionic strength are the same as in Fig.~\ref{Fig:potential}. The solid line is the curve $\langle x \rangle = 224.9 / \lambda$ which fits well the points described by Eq.~\ref{expansion} in this range of magnetic field.}
\label{Fig:var temp}
\end{figure}

\begin{figure}
{\par { \rotatebox{0}{\includegraphics[scale=1]{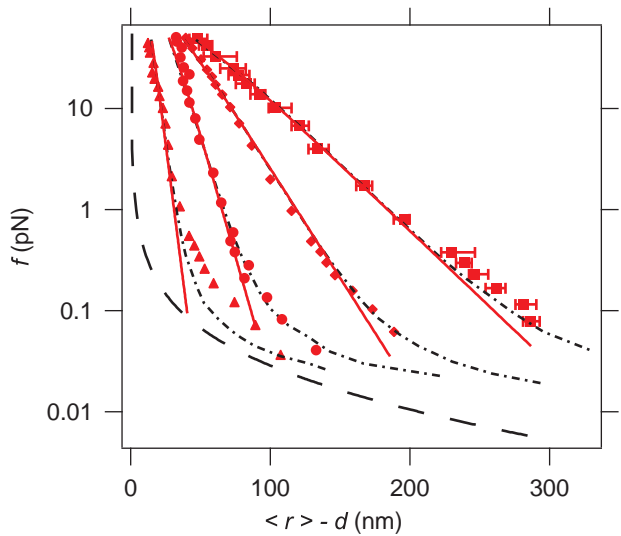}} }  
\par}
\caption{Force versus the mean distance between the bead surfaces $\langle r \rangle -d$
 for different ionic
strengths. Symbols correspond to experimental data points, from left to right, $\blacktriangle$
 5mM, $\bullet$ 1mM, $\blacklozenge$ 0.25mM and $\blacksquare$ 0.07mM.
The straight lines represent the electrostatic part of the force,
the dot-dashed lines represent Monte Carlo simulations, and the dashed line on the left represent the
expected force for a hard sphere potential.} \label{Fig:f_vs_d}
\end{figure}

We have carried out Monte-Carlo simulations of single chains,
which give direct access to the thermal average $\langle r
\rangle$. Each chain contains 10 to 20 particles interacting with
the potential $u(r)$. We have compared results of such simulations
to the low temperature expansion of Eq.~\ref{expansion}. As
expected, the low temperature approximation correctly describes
the simulations with the exact potential in the limit where the
effective temperature $1/\lambda$ is low. This is shown in
Fig.~\ref{Fig:var temp} for a particular choice of the ionic
strength and zeta potential. In the range of magnetic field
considered and for these conditions, the low temperature expansion
leads to a linear dependance of $\langle x \rangle$ as function of
the effective temperature with $\langle x \rangle \simeq 224.9 /
\lambda$ [nm]. Note that error bars on the simulation data of
figures \ref{Fig:var temp} and \ref{Fig:f_vs_d} were determined by
the method of Ref.~\cite{petersen}, they are very small everywhere
except for the smallest values of $\lambda$ near the limit of
stability of the chains.

From a practical point of view, an attractive feature of this
experiment is that it is possible to use it to measure a force
acting between neighboring particles. More precisely, this can be
done by defining the force $f(\langle r \rangle )$, from the
magnetic part of the potential as \bb f(\langle r \rangle
)=\left. \frac{du_{m}(r)}{dr} \right|_{r=\langle r \rangle }=\frac{3 \lambda k_B
T d^3}{\langle r \rangle ^4}, \en where $\langle r \rangle $ is
evaluated at the middle of the chain.

This choice of the middle of the chain is made to minimize finite
size effects, which we found from our numerical study have
 only a small effect on
the evaluation of forces. We have also found that forces
calculated assuming nearest neighbor interactions differ from the
case where all neighbors are included \cite{halsey}, only at high
magnetic field. In other words, taking into account long range
dipolar interactions does not change significantly the force
versus distance curves and thus is not essential for analyzing the
thermal expansion effect, as we have checked numerically.
Furthermore, by comparing the energy of the finite chain to that
of the infinite chain, we have found that the force versus
distance curves depend only weakly on the chain length, thus
confirming the Zhang and Widom model \cite{zhang95}, on which
Ref.~\cite{leal94} is based.

The force $f(\langle r \rangle )=f(r)$ defined in this way is
equal to the derivative of the electrostatic part of the potential
$f_{el}(r)=-du_{el}(r)/dr$ only when $1/\lambda=0$, {\it i.e.}
when $\langle x \rangle=0$. Thus a deviation of $f(r)$ away from
$f_{el}(r)$ can be attributed to thermal fluctuations.

As can be seen in fig.~\ref{Fig:potential}, at low values of
$\lambda$, the potential $u(r)$ is very steep for $r < r_{min}$
where it can be approximated by its short-ranged part
$u_{el}(r)+u_{HS}(r)$, while $u(r) \simeq u_m(r)$ for $r>r_{min}$.
The most asymmetric situation corresponds therefore to the limit
where the Debye length goes to zero. In this case, the magnetic
part of the potential is only balanced by the hard sphere
repulsion of the colloids. This case is more simple than the
previous case because now the position of the minimum of the
potential $r_{min}$ is always $d$ irrespective of $\lambda$. An
expansion of the potential about this minimum at $r=d$ leads to
\bb v(x)=px+ \frac{1}{2} kx^2, \en with $p=u'_m(d)=3 \lambda k_B
T/d$ and $k=u''_m(d)=-12 \lambda k_B T/d^2$. Now by treating the
term $kx^2/2$ as a perturbation with respect to the first term
$px$, one obtains to lowest order in $T$: $\langle x \rangle
\simeq k_B T/p$, and thus
\begin{equation}\label{scaling form of x}
\langle r \rangle  \simeq d \left( 1+\frac{1}{3\lambda} \right).
\end{equation}
This approximation holds for $\lambda \gg 2/3$, so essentially for
all values of the magnetic field considered here. At the same
order in perturbation, $\langle (x-\langle x \rangle )^2 \rangle  \simeq (k_B T/p)^2$, which
means that in this case the RMS fluctuations of $x$ are of the same
order than their average. For hard-sphere
potentials, Eq.~\ref{scaling form of x} can be inverted to obtain
$\lambda(\langle r \rangle )$ which leads to the following simple expression for
the force
\begin{equation}\label{HS scaling form of x}
f(\langle r \rangle )=\left( \frac{d}{\langle r \rangle } \right)^4 \frac{k_B T}{\langle r \rangle -d}.
\end{equation}
In the limit $T=0$, we recover that $f(\langle r \rangle )
\rightarrow 3 \mu_0 m^2/2\pi d^4$. The force $f(\langle r \rangle
)$ shown in Fig.~\ref{Fig:f_vs_d}, has a vertical
part near $\langle r \rangle=d$ and a slower power-law decay at
larger distances. The vertical part of the curve typical of hard
sphere behavior has been observed in Ref.~\cite{monval} but the
slower decay at larger distances has not been analyzed there
although it is expected to be present in these experiments.

Let us discuss measurements of the force versus distance \cite{remi}. We use Dynal-Invitrogen
magnetic beads of diameter $d \simeq 1150$nm, as observed by Dynamic Light Scattering.
The polydispersity in size is estimated to be in the range of $1\%$.
We vary the interaction potential by changing the
ionic strength, and we measure the mean particle distance using digital
video-microscopy \cite{grier}.
The mean distance is calculated by averaging the distance between particles
within the chain and over the time of the experiment.
The force versus the mean distance between bead surfaces
is shown
in Fig.~\ref{Fig:f_vs_d}.
Each set of experiment with a given ionic strength is performed
with the same chain in order to reduce errors due polydispersity in particle size.
To determine the electrostatic potential,
we have fitted the quasi-linear part of the curve for the zeta
potential $\Psi_0$, and we have found
$\Psi_0 \simeq -35$mV.
When the Debye length is large, the
expansion $\langle x \rangle$ is always negligible in the range of
force and distance which can be resolved. As a consequence, we observe only the low
temperature regime with a quasi-linear force (in log scale) versus distance
profile, characteristic of the electrostatic part of the interaction.
For smaller Debye length, thermal fluctuations make
the expansion $\langle x \rangle$ observable.
A very good agreement is found between the
simulations and the experiments except for the smallest Debye
length in the region where the deviation from the linear profile occurs.
We believe that this
discrepancy is not due to a failure of the electrostatic
description, which could only happen at much higher values of the
surface potential $\Psi_0$ than observed here. Instead we
attribute the discrepancy to an additional repulsive force, which
is due to a polymer layer grafted on the particles \cite{joanny}.
This force is indeed important only when the particles are
sufficiently close but irrelevant otherwise. When this additional
contribution is taken into account, a good agreement between the
experiments and the theory can be obtained for all the ionic
strengths studied here.

In Ref.~\cite{leal94}, a deviation in the force versus distance
measurements (with respect to a linear profile) was reported at high ionic strength and for
low field, which we attribute to the thermal expansion effect discussed in this paper.
The particles were a factor 5 smaller (there
$d=190$nm) than in the present study. In fact,
for smaller particles, a higher ionic strength is necessary to
observe the deviation of Fig.~\ref{Fig:f_vs_d}.
The importance of the particle size or of
the Debye length can be determined by evaluating the position of
the point where a deviation starts to become observable. If we assume that
this happens when $\langle x \rangle  \simeq 0.1 \times \langle r \rangle $, using
our low temperature approximation, we find that this point corresponds
to ($F=0.7$pN,$h=30$nm) for the curve at $c=5$mM and
($F=0.17$pN,$h=131$nm) for the curve at $c=0.25$mM in agreement
with Fig.~\ref{Fig:f_vs_d}. For lower ionic strength and lower forces,
the effect is too weak to be observable.

As the strength of the dipolar interaction is reduced, we observe
both in the experiments and in the simulations that thermal
fluctuations cause chains to fragment. Independent of
fragmentation, disordering begins at the ends \cite{philipps},
where the inter-particle distance is larger due to finite size
effects. This effect is similar to edge and surface melting in 2D
and 3D solids. We have studied numerically the ratio of the
standard deviation of the inter-particle distance divided by the
average inter-particle distance, $L_e=\sqrt{\langle x^2 \rangle
}/\langle r \rangle $, which is similar to the Lindemann parameter
used to detect the approach to melting in 3D systems. At large
$\lambda$, the potential is harmonic and $L_e$ is small, whereas
for small $\lambda$, the potential becomes anharmonic and $L_e$
increases. We find that $L_e$ raises very steeply as it approaches
a value of the order of 0.1, which corresponds approximatively to
melting in 3D systems. The threshold value on $\lambda$ where this
happens depends very much on dimension, it is much lower in 2D as
in 1D for instance.

In order to investigate the 2D case, we have developed a 2D
extension of our Monte Carlo simulation, in which the location of
the magnetic dipoles of the chain form a 2D polymer. In addition,
we have assumed that the orientations of these dipoles remain
frozen in the direction of the applied magnetic field. We have
found that the force versus distance curves shown in
Fig.~\ref{Fig:f_vs_d} do not differ in the 2D case as compared
with the 1D case, in the range of magnetic field considered here.
A difference between the predictions of the 2D and 1D model only
arise at very small forces and large distances. In the 2D case,
transverse modes of fluctuations of the chain arise, which are
soft bending modes \cite{toussaint}. Such soft modes do not
significantly alter the mean distance, which is represented in
Fig.~\ref{Fig:f_vs_d}. This explains the robustness of the 1D
model. The region where thermal fluctuations are very large
requires further modelling, since there the 2D or 3D character of
the problem is important.

To summarize, we have analyzed the thermal expansion of a chain of
magnetic particles. We have shown that this effect is responsible
for a deviation with respect to the quasi-linear force profile
observed at low effective temperatures. This effect can be
described using a simple 1D model, and is relevant to the
determination of force versus distance using magnetic colloids.

We acknowledge fruitful discussions with P. Chaikin, JF. Joanny, R. Dreyfus and F. Krzakala. D. L.
also acknowledges support from the Indo-French Center CEFIPRA (grant 3504-2).

\bibliographystyle{apsrev}
\bibliography{force}
\end{document}